# QSMDiff: Unsupervised 3D Diffusion Models for Quantitative Susceptibility Mapping

Zhuang Xiong[1], Wei Jiang[1], Yang Gao[1,2], Feng Liu[1], Hongfu Sun[1,3]

1. School of Electrical Engineering and Computer Science, the University of Queensland, Australia
2. School of Computer Science and Engineering, Central South University, China
3. School of Engineering, University of Newcastle, Australia

**Abstract.** Quantitative Susceptibility Mapping (QSM) dipole inversion is an ill-posed inverse problem for quantifying magnetic susceptibility distributions from MRI tissue phases. While supervised deep learning methods have shown success in specific QSM tasks, their generalizability across different acquisition scenarios remains constrained. Recent developments in diffusion models have demonstrated potential for solving 2D medical imaging inverse problems. However, their application to 3D modalities, such as QSM, remains challenging due to high computational demands. In this work, we developed a 3D image patch-based diffusion model, namely QSMDiff, for robust QSM reconstruction across different scan parameters, alongside simultaneous super-resolution and image-denoising tasks. QSMDiff adopts unsupervised 3D image patch training and full-size measurement guidance during inference for controlled image generation. Evaluation on simulated and in-vivo human brains, using gradient-echo and echo-planar imaging sequences across different acquisition parameters, demonstrates superior performance. The method proposed in QSMDiff also holds promise for impacting other 3D medical imaging applications beyond QSM.

## 1  Introduction

Quantitative Susceptibility Mapping (QSM) is a Magnetic Resonance Imaging (MRI) post-processing technique for measuring tissue magnetic susceptibility, which has applications in various neurological diseases [1-7]. However, dipole inversion, a key step in the QSM processing framework that involves a deconvolution operation with the magnetic dipole kernel, is intrinsically an ill-posed inverse problem and would be even more challenging for accelerated acquisitions such as Echo-Planar Imaging (EPI) [8, 9] due to the poor image resolution and signal-to-noise ratio (SNR).

An increasing number of deep learning-based dipole inversion methods [10-12] have been developed using supervised approaches. However, despite recent efforts in enhancing their robustness through strategies such as physics guidance [13-15] and acquisition-specific refinement [16, 17], these methods continue to exhibit limited model generalizability across varying image resolutions, SNRs, and acquisition orientations.

Generative models employing generative adversarial networks (GANs) and variational auto-encoders (VAEs) have seen widespread use in synthesizing both natural and medical images. However, the generative processes of GANs and VAEs are characterized by limited controllability, rendering them less reliable for medical image reconstruction tasks required conformity with instrumental measurements. Recently, diffusion models [18, 19] and score-based generative models [20] have gained significant attention thanks to their excellent capability of controllable high-quality image generation in a zero-shot setting. Unfortunately, these models demand significantly greater computational resources, which generally surpass the capacity of current hardware to perform full-size 3D image reconstruction tasks.

In this study, we introduce QSMDiff, a 3D image patch-based diffusion model designed for robust QSM dipole inversion from various acquisition scenarios. It notably supports simultaneous super-resolution and image denoising for EPI-QSM acquisitions, highlighting its efficacy and versatility. QSMDiff significantly improves the feasibility of diffusion models, resulting in considerable memory savings. Furthermore, it preserves the accuracy and quality of image generation consistent with measurement conditions. Inspired by prior work [21] in 2D, we develop an overlapping cropping mechanism tailored for arbitrary matrix sizes in the 3D scenario to address the discontinuity and boundary artifact arising from patch generation. Our extensive experiments and quantitative evaluations demonstrate that QSMDiff excels in superior model generalizability compared with several state-of-the-art QSM methods and achieves high-quality QSM dipole inversion across diverse scan parameters such as different image resolutions and acquisition orientations.

## 2    Method

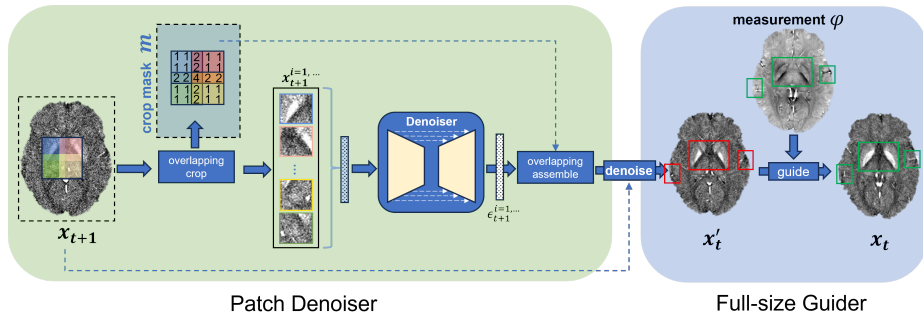

**Fig. 1.** Illustration of a sampling step for a full-size 3D volume during inference, featuring a Patch Denoiser for unconditional sampling and a Full-size Guider for measurement-conditioned sampling. Structural corrections, marked in red, demonstrate the Guider's ability to accurately reconstruct the true structure, aligned with measurements indicated by green rectangles.

### 2.1 Patch-based diffusion model

The QSMDiff framework introduces an innovative approach by integrating an unsupervised diffusion model as a generative prior with a conditional sampling process suited for various measurements. In the training phase, explained in Algorithm 1, full-size 3D QSM volumes $x_0$ are cropped into smaller, overlapping patches $x_0^i$. Following the Denoising Diffusion Probabilistic Models (DDPM) approach [18], this unsupervised training includes a forward phase that gradually adds noise $\epsilon_t^i$ to the image patches and a backward phase that aims to recover the original patches by predicting the noise added to its disturbed versions $x_t^i$. The training loss is defined as:

$$L(\theta) = E_{x,\epsilon,t,i}[||\epsilon_t^i - \epsilon_\theta^i(x_t^i, t)||^2], \quad (1)$$

where $\epsilon_\theta^i$ symbolizes the noise prediction from the parameterized diffusion model, given a disturbed patch $x_t^i$ and the time step $t$ as inputs.

### 2.2 Conditional sampling for QSM dipole inversion

During the inference phase, as depicted in Fig. 1 and elaborated in Algorithm 2, the full-size 3D volume $x_t$ is cropped into overlapping patches $x_t^i$ by setting patch size $d$ and overlap size $o$ in the Patch Denoiser. A crop mask is generated to record the overlap frequency for each voxel. The noise predicted for each patch $\epsilon_\theta^i(x_t^i, t)$ is then reassembled based on their original positions and normalized using the crop mask. The unconditional denoising step for the full-size volume is then completed by substracting the assembled noise from the noisy full-size image $x_t$.

In the subsequent Full-size Guider, conditional sampling is performed under the guidance of a local field map preprocessed from the gradient-echo phase measurement. The relationship between field ($\varphi$) and susceptibility ($\chi$) can be expressed as a piecewise point multiplication in the Fourier domain:

$$\varphi = \mathcal{F}^{-1}D\mathcal{F}\chi + \eta, \quad (2)$$

where $\mathcal{F}$ and $\mathcal{F}^{-1}$ denote the forward and inverse Fourier transforms, $\eta$ is measurement noise, and $D$ is the unit dipole kernel.

We adopt the Diffusion Posterior Sampling [22] (DPS) strategy to solve the dipole inversion problem. As described in Algorithm 2, after the unconditional denoising step, which yielded $x_{t-1}$, the QSMDiff conditional sampling step can be formulated as:

$$\bar{x}_{t-1} = x_{t-1} - \xi_1 \cdot \nabla_{x_t}\text{DipInv}(\varphi, \hat{x}_0) - \xi_2 \cdot \nabla_{x_t}\text{Trans}(x_{tkd}, \hat{x}_0) - \lambda \cdot \nabla_{x_t}\text{TV}(\hat{x}_0), \quad (3)$$

where DipInv is the model loss:

$$\text{DipInv}(\varphi, \hat{x}_0) = ||\varphi - \mathcal{F}^{-1}D\mathcal{F}S\hat{x}_0||^2, \quad (4)$$

and Trans is the image-to-image translation loss:

$$\text{Trans}(x_{tkd}, \hat{x}_0) = ||x_{tkd} - S\hat{x}_0||^2, \quad (5)$$

with $\hat{x}_0$ denotes the predicted posterior mean of the full-size QSM and $x_{tkd}$ denotes the initial QSM estimate from the Thresholded K-space Division (TKD) method [23]. $S$ is the image resampling operation when there is an image resolution mismatch between measurement and pre-trained diffusion prior. $\xi_1$ and $\xi_2$ are weighting factors balancing the guidance contribution. TV denotes for total variation penalty that preserves image smoothness, ensuring the spatial continuity of the patch-assembled full brain volume.

| **Algorithm 1.**   Training | **Algorithm 2.**   Sampling |
|---|---|
| **repeat** | $x_T \sim N(0, I)$, **select** $d, o, \xi_1, \xi_2, \lambda$ |
| $x_0^i \sim q(\{x_0^1, x_0^2, \dots, x_0^n\})$ | **for** $t = T, \dots, 1$ **do** |
| $t \sim \{1, \dots, T\}$ | $\sum_{i=0}^{n} x_t^i, m = P(x_t, d, o)$ |
| $\epsilon \sim N(0, I)$ | $z \sim N(0, I), \sigma_t \sim \{\sigma_{t=1:T}\}$ |
| $\bar{\alpha}_t = \prod_{k=1}^{t} \alpha_k \,;\, \alpha_t = 1 - \beta_t$ | $\epsilon = P^{-1}(\epsilon_\theta \left( \sum_{i=0}^{n} x_t^i, t \right), d, o)/m$ |
| $x_t^i = \sqrt{\bar{\alpha}_t} x_0^i + \sqrt{1 - \bar{\alpha}_t}\epsilon$ | $x_{t-1} = \frac{1}{\sqrt{\alpha_t}}(x_t - \frac{(1 - \alpha_t)}{\sqrt{(1 - \bar{\alpha}_t)}}\epsilon) + \sigma_t z$ |
| **gradient descent on:** | $\hat{x}_0 = \frac{1}{\sqrt{\bar{\alpha}_t}}(x_t - \sqrt{1 - \bar{\alpha}_t}\epsilon)$ |
| $\nabla_\theta \|\epsilon - \epsilon_\theta(x_t^i, t)\|^2$ | $\bar{x}_{t-1} = x_{t-1} - \xi_1 \cdot \nabla_{x_t}\text{DipInv}(\varphi, \hat{x}_0) - \xi_2 \cdot \nabla_{x_t}\text{Trans}(x_{tkd}, \hat{x}_0) - \lambda \cdot \nabla_{x_t}\text{TV}(\hat{x}_0)$ |
| **until** converged | **end for** |
| | **return** $\hat{x}_0$ |

## 3       Experiments

QSMDiff was trained based on two open-public COSMOS [24] datasets with an isotropic image resolution of 1 mm [13, 25], from which a total of 18,000 three-dimensional patches of size $48^3$ were then generated with a stride of 32 at each dimension. Patches containing over 95% zero-value voxels were excluded from the training dataset. Values in all QSM patches were normalized to the range [-1, -1]. Trilinear interpolation was employed here as the resolution resampling operator $S$.

To enhance the computational efficiency, the overlap size $o$ was experimentally set to 8 for each dimension during sampling. The TKD threshold for QSM initial estimate was set to 0.1. We empirically tuned the weight factor $\xi_1$ and $\xi_2$ to be 10 and 2.5, respectively, which demonstrated on average optimal results in different scenarios.

The code implementation of QSMDiff was modified from the Ablated Diffusion Model [19] with a classic 3D UNet structure. Denoising Diffusion Implicit Models sampling [26] with 200 steps was adopted to speed up the sampling process. All the work was developed using Python 3.10 and Pytorch 2.0. QSMDiff was trained for 5

Robust quantitative susceptibility mapping using 3D Diffusion models days using one Nvidia H100 GPU with 80GB vRAM, and the sampling was conducted using a RTX 4090 GPU with 24GB vRAM.

### 3.1 Ablation Study

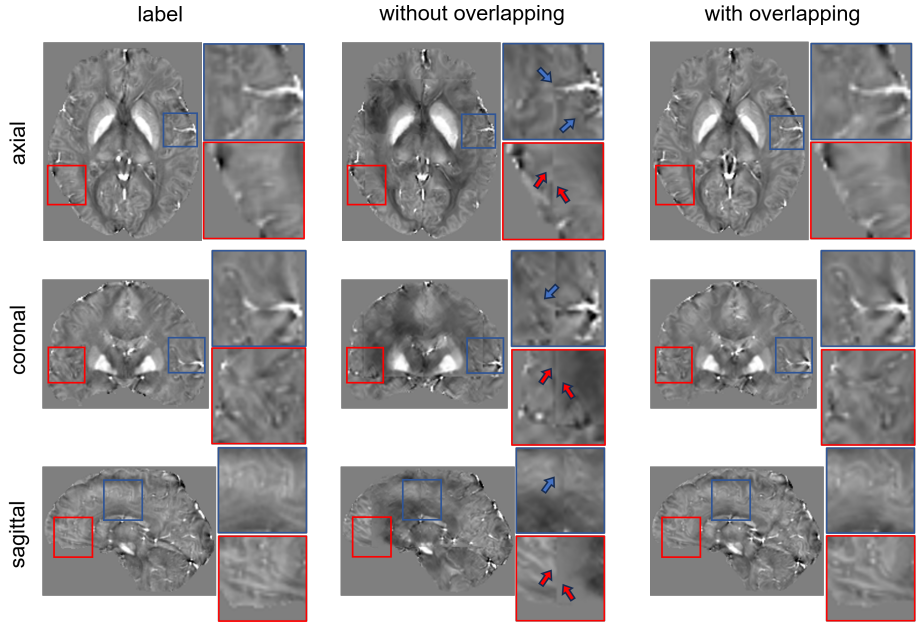

**Fig. 2.** Comparisons of QSMDiff with and without the overlapping patch training strategy. Arrows point to boundary artifacts and structural inconsistencies. Color range: [-0.15, 0.15] ppm.

An ablation study was performed on a 1 mm$^3$ COSMOS simulated human brain with pure-axial head orientation to evaluate the effectiveness of the proposed overlapped-cropping strategy. As shown in Fig. 2, QSM results generated without overlapping patches exhibited shadow and boundary artifacts, as well as structural inconsistencies, as pointed by the red and blue arrows within the zoomed-in regions. In contrast, the proposed algorithms with overlapped cropping successfully suppressed the artifacts and preserved more fine structural details.

### 3.2 GRE dipole inversion test

QSMDiff is compared with several established QSM methods (including two iterative methods (iLSQR [27] and MEDI [28]) and three deep learning methods (Unet trained on 1 mm isotropic and pure-axial dataset, AFTER-QSM [15] and LPCNN [13] trained by their original authors)) on an anisotropic resolution (vox= [1 × 1 × 3] mm$^3$) dataset. As shown in Fig. 3, QSMDiff demonstrated the most appealing results, maintaining excellent susceptibility contrast. Notably, artifacts associated with parallel imaging, highlighted by red rectangles in the in-vivo local field images, were effectively

6       Zhuang Xiong et al.

removed in the QSMDiff results, but persisted in other methods. Moreover, as reported in Fig. 4(a) and Table 1, QSMDiff exhibited visually and numerically the best performance on a simulated human brain with an anisotropic resolution (vox= $[1 \times 1 \times 3]$ mm³) and a titled head orientation ($\vec{p} = [0.5, 0.5, 0.71]$).

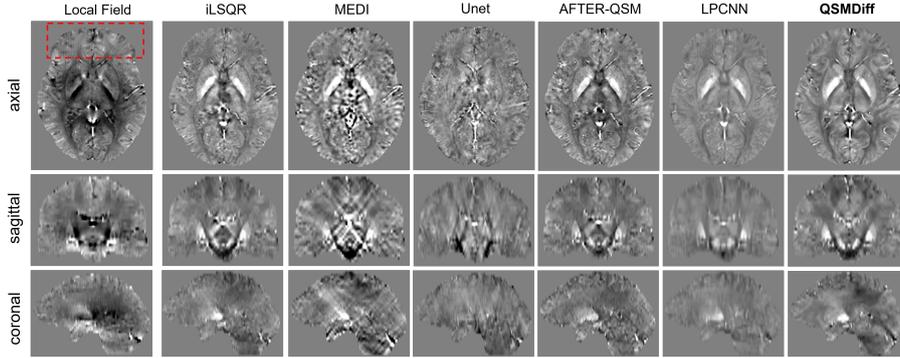

**Fig. 3.** Method comparison on an in-vivo brain with $[1 \times 1 \times 3]$ mm³ resolution in pure-axial head orientation. The red rectangle highlights the parallel imaging artifact. Sagittal and coronal views were adjusted to 1 mm isotropic for clearer visualization. Color range: [-0.15, 0.15] ppm.

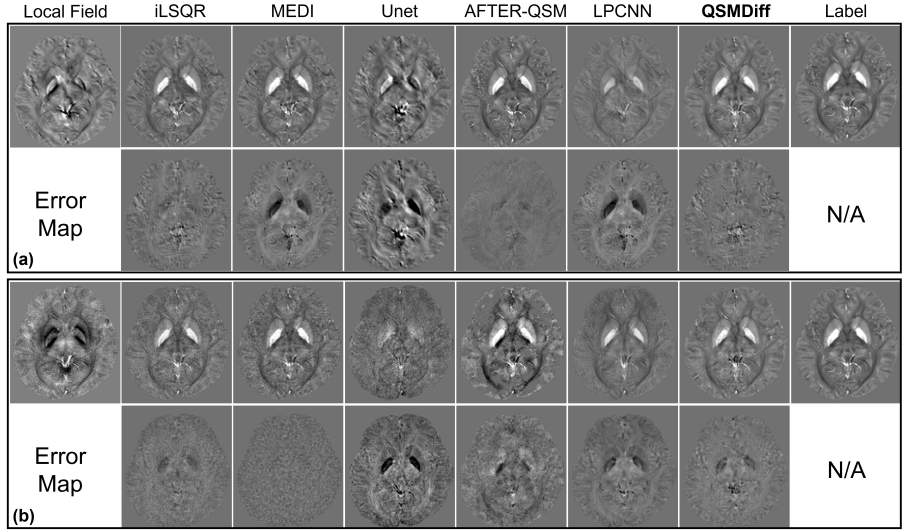

**Fig. 4.** Visual comparisons between QSMDiff and other methods on simulated datasets with (a) tilted acquisition orientation ($\vec{p} = [0.5, 0.5, 0.71]$), and (b) noisy local field measurement. Color range: [-0.12, 0.15] ppm.

### 3.3  Simultaneous super-resolution & dipole inversion test

In Fig. 5, the performance of QSMDiff was further evaluated through a simultaneous super-resolution & dipole inversion test, using COSMOS simulated GRE local fields

Robust quantitative susceptibility mapping using 3D Diffusion models at different low image resolutions The method successfully maintained image contrast across all orthogonal views and resolutions. Notably, it excelled in recovering fine details in dipole inversions from lower resolutions up to 2 mm isotropic. QSMDiff also performed remarkably well at an extremely low resolution of 2×2×4 mm³, effectively reducing blurriness and introducing only minor structural discrepancies that were not consistent with the 1 mm³ resolution label.

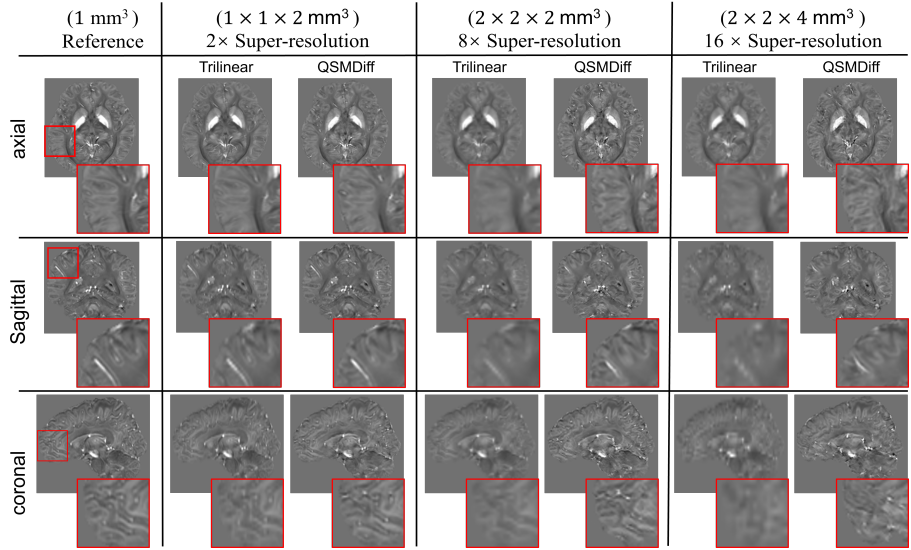

**Fig. 5.** QSMDiff results on 3 simulated low-resolution GRE local field maps. For each resolution case, the first column shows results from trilinear interpolation, while the second column displays reconstructions from QSMDiff. Color range: [-0.12, 0.15] ppm.

### 3.4 Simultaneous denoising & dipole inversion test

**Table 1.** Performance metrics on simulated titled and noisy human brain datasets.

| Method | PSNR | | SSIM | | HFEN | |
|---|---|---|---|---|---|---|
| | Titled | Noisy | Titled | Noisy | Titled | Noisy |
| iLSQR | 36.34 | 36.78 | 0.819 | 0.829 | 41.23 | 52.29 |
| MEDI | 36.65 | 37.34 | 0.842 | 0.823 | 42.39 | 49.82 |
| Unet | 30.91 | 31.31 | 0.711 | 0.684 | 57.88 | 72.66 |
| AFTER-QSM | 41.10 | 33.91 | 0.912 | 0.784 | **20.92** | 55.93 |
| LPCNN | 35.36 | 36.41 | 0.788 | 0.812 | 37.01 | 34.00 |
| **QSMDiff** | **41.16** | **37.79** | **0.913** | **0.853** | 21.64 | **32.32** |

The robustness of QSMDiff against noise corruption was visually assessed in Fig. 4(b) on the same COSMO human brain but was corrupted with noise. The corresponding Peak Signal-to-Noise Ratio (PSNR), Structural Similarity index Measure (SSIM) and High Frequency Error Norm (HFEN) were reported in Table 1. QSMDiff achieved the best reconstruction accuracy over iterative and other deep learning methods.

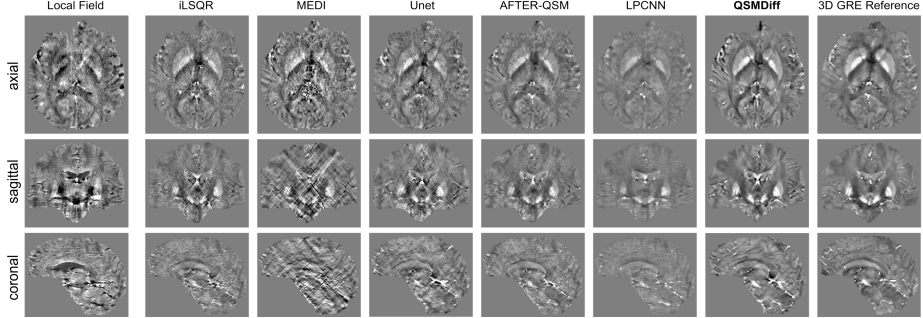

**Fig. 6.** Comparisons of various QSM methods on an in-vivo 2D EPI acquisition feature a 1 mm³ image resolution but low SNR in a pure-axial orientation. Color range: [-0.15, 0.15] ppm.

In Fig. 6, QSMDiff was compared to existing methods employing an in-vivo 2D EPI acquisition with substantially lower SNR as shown in the first column. It is evident that QSMDiff achieved visually the most robust results compared to the other approaches in terms of noise removal, and structural detail preservation. AFTER-QSM and iLSQR showed the second-best performance but with apparent residual noise persisted. LPCNN, while producing a relatively less noisy output, resulted in underestimated reconstruction with respective to the GRE reference. MEDI displayed severe noise and streaking artifacts. This test underscores QSMDiff's potential in translating EPI QSM into GRE COSMOS style susceptibility maps, which are characterized by rich structural details and enhanced SNR and image contrast.

## 4    Discussion and conclusion

In this work, we proposed a novel image patch-based 3D diffusion model, QSMDiff, for the challenging QSM dipole inversion problem. QSMDiff has shown remarkable generalizability across different acquisition parameters and has outperformed existing state-of-the-art methods in terms of its robustness against artifacts, noise, and variations in acquisition parameters. It also proved capable of significantly reducing QSM scan times through ultra-fast EPI acquisitions without substantially sacrificing image quality and accuracy. Leveraging patch training and full-size measurement guidance techniques, QSMDiff effectively addresses the difficulties presented by high-resolution 3D QSM for diffusion models. However, this study recognizes several limitations, including the untested applicability of QSMDiff on diseased brains. Furthermore, due to its iterative sampling process, QSMDiff is more time-consuming compared to current QSM methods. Future efforts will focus on extending QSMDiff's application to brains with abnormalities, such as tumors and hemorrhages, and on speeding up its conditional sampling process through more advanced techniques.

Robust quantitative susceptibility mapping using 3D Diffusion models

# References

1. Acosta-Cabronero, J., Williams, G.B., Cardenas-Blanco, A., Arnold, R.J., Lupson, V., Nestor, P.J.: In vivo quantitative susceptibility mapping (QSM) in Alzheimer's disease. PLoS One 8, e81093 (2013)
2. Chen, W., Zhu, W., Kovanlikaya, I., Kovanlikaya, A., Liu, T., Wang, S., Salustri, C., Wang, Y.: Intracranial calcifications and hemorrhages: characterization with quantitative susceptibility mapping. Radiology 270, 496-505 (2014)
3. Elkady, A.M., Cobzas, D., Sun, H., Seres, P., Blevins, G., Wilman, A.H.: Five year iron changes in relapsing-remitting multiple sclerosis deep gray matter compared to healthy controls. Mult Scler Relat Disord 33, 107-115 (2019)
4. Balla, D.Z., Sanchez-Panchuelo, R.M., Wharton, S.J., Hagberg, G.E., Scheffler, K., Francis, S.T., Bowtell, R.: Functional quantitative susceptibility mapping (fQSM). Neuroimage 100, 112-124 (2014)
5. Sun, H., Seres, P., Wilman, A.: Structural and functional quantitative susceptibility mapping from standard fMRI studies. NMR in Biomedicine 30, e3619 (2017)
6. Acosta-Cabronero, J., Cardenas-Blanco, A., Betts, M.J., Butryn, M., Valdes-Herrera, J.P., Galazky, I., Nestor, P.J.: The whole-brain pattern of magnetic susceptibility perturbations in Parkinson's disease. Brain 140, 118-131 (2017)
7. van Bergen, J.M., Hua, J., Unschuld, P.G., Lim, I.A., Jones, C.K., Margolis, R.L., Ross, C.A., van Zijl, P.C., Li, X.: Quantitative Susceptibility Mapping Suggests Altered Brain Iron in Premanifest Huntington Disease. AJNR Am J Neuroradiol 37, 789-796 (2016)
8. Sun, H., Wilman, A.H.: Quantitative susceptibility mapping using single-shot echo-planar imaging. Magn Reson Med 73, 1932-1938 (2015)
9. Langkammer, C., Bredies, K., Poser, B.A., Barth, M., Reishofer, G., Fan, A.P., Bilgic, B., Fazekas, F., Mainero, C., Ropele, S.: Fast quantitative susceptibility mapping using 3D EPI and total generalized variation. Neuroimage 111, 622-630 (2015)
10. Yoon, J., Gong, E., Chatnuntawech, I., Bilgic, B., Lee, J., Jung, W., Ko, J., Jung, H., Setsompop, K., Zaharchuk, G., Kim, E.Y., Pauly, J., Lee, J.: Quantitative susceptibility mapping using deep neural network: QSMnet. Neuroimage 179, 199-206 (2018)
11. Bollmann, S., Rasmussen, K.G.B., Kristensen, M., Blendal, R.G., Ostergaard, L.R., Plocharski, M., O'Brien, K., Langkammer, C., Janke, A., Barth, M.: DeepQSM - using deep learning to solve the dipole inversion for quantitative susceptibility mapping. Neuroimage 195, 373-383 (2019)
12. Gao, Y., Zhu, X., Moffat, B.A., Glarin, R., Wilman, A.H., Pike, G.B., Crozier, S., Liu, F., Sun, H.: xQSM: quantitative susceptibility mapping with octave convolutional and noise-regularized neural networks. NMR Biomed 34, e4461 (2021)
13. Lai, K.W., Aggarwal, M., van Zijl, P., Li, X., Sulam, J.: Learned Proximal Networks for Quantitative Susceptibility Mapping. Med Image Comput Comput Assist Interv 12262, 125-135 (2020)
14. Feng, R., Zhao, J., Wang, H., Yang, B., Feng, J., Shi, Y., Zhang, M., Liu, C., Zhang, Y., Zhuang, J., Wei, H.: MoDL-QSM: Model-based deep learning for quantitative susceptibility mapping. Neuroimage 240, 118376 (2021)
15. Zhuang, X., Yang, G., Feng, L., Hongfu, S.: Affine transformation edited and refined deep neural network for quantitative susceptibility mapping. NeuroImage 267, 119842 (2023)

16. Xiong, Z., Gao, Y., Liu, Y., Fazlollahi, A., Nestor, P., Liu, F., Sun, H.: Quantitative Susceptibility Mapping through Model-based Deep Image Prior (MoDIP). arXiv preprint arXiv:2308.09467 (2023)
17. Zhang, J., Liu, Z., Zhang, S., Zhang, H., Spincemaille, P., Nguyen, T.D., Sabuncu, M.R., Wang, Y.: Fidelity imposed network edit (FINE) for solving ill-posed image reconstruction. Neuroimage 211, 116579 (2020)
18. Ho, J., Jain, A., Abbeel, P.: Denoising diffusion probabilistic models. Advances in neural information processing systems 33, 6840-6851 (2020)
19. Dhariwal, P., Nichol, A.: Diffusion models beat gans on image synthesis. Advances in neural information processing systems 34, 8780-8794 (2021)
20. Song, Y., Sohl-Dickstein, J., Kingma, D.P., Kumar, A., Ermon, S., Poole, B.: Score-based generative modeling through stochastic differential equations. arXiv preprint arXiv:2011.13456 (2020)
21. Özdenizci, O., Legenstein, R.: Restoring vision in adverse weather conditions with patch-based denoising diffusion models. IEEE Transactions on Pattern Analysis and Machine Intelligence (2023)
22. Chung, H., Kim, J., Mccann, M.T., Klasky, M.L., Ye, J.C.: Diffusion posterior sampling for general noisy inverse problems. arXiv preprint arXiv:2209.14687 (2022)
23. Wharton, S., Schäfer, A., Bowtell, R.: Susceptibility mapping in the human brain using threshold-based k-space division. Magnetic resonance in medicine 63, 1292-1304 (2010)
24. Liu, T., Spincemaille, P., De Rochefort, L., Kressler, B., Wang, Y.: Calculation of susceptibility through multiple orientation sampling (COSMOS): a method for conditioning the inverse problem from measured magnetic field map to susceptibility source image in MRI. Magnetic Resonance in Medicine: An Official Journal of the International Society for Magnetic Resonance in Medicine 61, 196-204 (2009)
25. Shi, Y., Feng, R., Li, Z., Zhuang, J., Zhang, Y., Wei, H.: Towards in vivo ground truth susceptibility for single-orientation deep learning QSM: A multi-orientation gradient-echo MRI dataset. NeuroImage 261, 119522 (2022)
26. Song, J., Meng, C., Ermon, S.: Denoising diffusion implicit models. arXiv preprint arXiv:2010.02502 (2020)
27. Li, W., Wang, N., Yu, F., Han, H., Cao, W., Romero, R., Tantiwongkosi, B., Duong, T.Q., Liu, C.: A method for estimating and removing streaking artifacts in quantitative susceptibility mapping. Neuroimage 108, 111-122 (2015)
28. Liu, T., Liu, J., de Rochefort, L., Spincemaille, P., Khalidov, I., Ledoux, J.R., Wang, Y.: Morphology enabled dipole inversion (MEDI) from a single-angle acquisition: comparison with COSMOS in human brain imaging. Magn Reson Med 66, 777-783 (2011)